# Tunable Angle Dependent Magnetization Dynamics in $Ni_{80}Fe_{20}$ Nano-cross Structures of Varying Size

Kartik Adhikari[1], Saswati Barman[2], Ruma Mandal[1], Yoshichika Otani[3,4] and Anjan Barman[1,*]

[1]*Department of Condensed Matter Physics and Material Sciences, S. N. Bose National Centre for Basic Sciences, Block JD, Sector III, Salt Lake, Kolkata 700106, India*

[2]*Institute of Engineering and Management, Sector V, Salt Lake, Kolkata 700091, India*

[3]*CEMS-RIKEN, 2-1 Hirosawa, Wako, Saitama 351-0198, Japan*

[4]*Institute for Solid State Physics, University of Tokyo, 5-1-5 Kashiwanoha, Kashiwa, Chiba 277-8581, Japan*

*\*Email address: abarman@bose.res.in*

## ABSTRACT

We demonstrate a large angular dependence of magnetization dynamics in $Ni_{80}Fe_{20}$ nano-cross arrays of varying sizes. By subtle variation of the azimuthal angle ($\phi$) of an in-plane bias magnetic field, the spin configuration and the ensuing spin-wave dynamics, including mode softening, mode splitting, mode crossover and mode merging, can be drastically varied to the extent that a frequency minimum corresponding to mode softening converts to a mode crossover, various mode splitting and mode crossover disappear and additional mode splitting appears. Numerically simulated spin-wave spectra and phase profiles revealed the nature of various spin-wave modes and the origin of above variation of the dynamics with bias-field angle. All of these above observations are further modified with the variation of dimensions of the nano-cross. The numerically calculated magnetostatic field distributions further supports the variation of dynamics with bias-field angle. Theseresults open a new avenue for engineering the nano-cross based magnetic devices such as magnetic storage, spin-wave logic and on-chip data communication devices.

# I. INTRODUCTION

Arrays of nanoscale ferromagnetic dots [1] have attracted ample interest in recent years, both in terms of fundamental physics as well as for their potential applications. During the last one decade, investigation of spin dynamics and spin wave (SW) [2] in such nanostructures has emerged as a very potent research area. This interest is triggered by the possibility of exploring new physics in such structures. On the other hand, due to the progress of nanolithography [3], patterned nanostructures and their arrays have been emerged as systems having great potential applications such as magnetic data storage [4,5], memory [6], logic devices [7] and spin torque nano-oscillators [8]. In nanodot arrays [9], the high surface to volume ratio, inhomogeneous demagnetizing field, dipole-dipole [10,11] interaction between the nanodots have significant effects on their magnetic properties and can lead to complex spin configuration within a single nanomagnet [12,13] and arrays of nanomagnets [14-20]. These complex spin configurations lead to rich variety of SW modes which have strong dependence on the strength [15] and orientation [16,17] of the applied bias magnetic field, and on the shape [18-20] and size [9,33] of nano-elements. Extensive research works have been carried out to investigate the magnetization dynamics in two dimensional (2-D) arrays of nanomagnets both experimentally using time-resolved scanning Kerr microscopy (TRSKM) [21] or time-resolved magneto-optical Kerr effect (TR-MOKE) microscopy [22], ferromagnetic resonance (FMR) [23-25], and Brillouin light scattering (BLS) [26,27]; and theoretically by micromagnetic simulations [28] and other numerical and analytical methods. While the optical techniques can extract the information about local behaviour of the dynamics, the ferromagnetic resonance can measure the global dynamics of a large array. In addition, due to its advantage of being a much faster measurement technique, very detailed investigation of bias field strength and angle dependence study can be made using this technique.

Ferromagnetic cross structure received some interest in the magnetism community due to its complex spin configuration [29]. However, the ultrafast magnetization dynamics measured by time-resolved magneto-optical Kerr effect revealed very rich dynamics with a strong configurational anisotropy [30]. Later in 2015, a report [31] proposed that ferromagnetic cross-shaped elements can be used as reconfigurable spin-based logic device using SW scattering and interference. More recent study [32] on cross-shaped nanodot arrays using broadband ferromagnetic resonance (FMR) measurements showed a bias field tunable magnetic configuration and magnetization dynamics, including the presence of mode softening and mode crossover. The above results opened up the possibility of application of ferromagnetic nano-cross structures as a potential building block of magnetic storage, memory, on-chip data communication, and spin-based logic devices, and hence, thorough investigation of the magnetization dynamics of this structure with its geometric parameters and external magnetic fields has become imperative. Here, we present the experimental and numerical study of the tunability of magnetization dynamics in $Ni_{80}Fe_{20}$ (permalloy, Py hereafter) nano-crosses of varying arm lengths (L) with in-plane orientation of the bias magnetic field. We use broadband ferromagnetic resonance technique and micromagnetic simulations for this work. We show that SW mode softening can be efficiently tuned by a subtle variation of the bias magnetic field orientation. Further properties such as SW mode splitting, mode crossover, mode merging and number of modes can also be easily tuned by bias field orientation, which further depend significantly on the size of the nano-cross element. We finally discuss some possible applications of the nano-cross arrays based on our observations.

## II. EXPERIMENTAL DETAILS

Arrays (200 µm × 20 µm) of Py nano-crosses with arm length (L) varying between 400 nm ≤ L ≤ 600 nm, and constant edge-to-edge separation (S) of 150 nm and thickness of 20 nm, as well as a continuous Py film of 20 nm thickness were fabricated on self-oxidized Si-substrate (001) by a combination of e-beam lithography and e-beam evaporation. Py film of 20 nm thickness coated with a 60-nm-thick $Al_2O_3$ protective layer was deposited in an ultra-high vacuum chamber at a base pressure of $2 \times 10^{-8}$ Torr on a bi-layer (PMMA/MMA) resist pattern on the Si substrate made by using e-beam lithography. A co-planer waveguide (CPW) made by Au of 150 nm thickness, 30 µm central conductor width, 300 µm length and 50 Ω nominal characteristic impedance was deposited on top of the nano-cross structures and the continuous Py film at a base pressure of $6 \times 10^{-7}$ Torr. Subsequently, a Ti protective layer of 5 nm thickness was deposited on top of the Au layer at the same base pressure. The waveguide was patterned by using mask-less photolithography. The broadband FMR experiments were performed using a vector network analyzer (Agilent, PNA-L N5230C, 10 MHz to 50 GHz) and a homebuilt high frequency probe station with non-magnetic G-S-G type probe (GGB Industries, Model No. 40A-GSG-150-EDP) [33]. A microwave output excitation is swept in broad frequency range with power of -15 dBm and fed into the CPW structure, generating a microwave magnetic field $h_{rf}$ along the y-axis of the nano-cross array. Additionally, an in-plane magnetic field, H, is applied along varying in-plane angle ϕ with respect to the x-axis and the output signal ($S_{11}$) is collected from the CPW in the reflection geometry. The measured reflection spectra are normalized by a reference measurement at high static magnetic field. A rotating electromagnet is used to apply an in-plane bias magnetic field up to 1.6 kOe. All the experiments are carried out at room temperature.

## III. RESULTS AND DISCUSSION

Figure 1(a) shows a schematic of the experimental set up. Figures 1(b-d) show scanning electron micrographs (SEMs) of all three arrays with cross-arm length (L) of 600, 500 and 400 nm. The bias magnetic field orientation ($\phi$) is shown in Fig. 1(b). The SEM images show that the cross structures suffer from some edge deformations and rounded corners, which increase with the reduction of L. The sizes of the individual crosses and their separations in the arrays also vary by up to about $\pm 10\%$. The external bias field (H) dependent FMR frequency (f) of the Py thin film is also measured and the data is fitted with the Kittel formula [34], which is given by:

$$f = \frac{\gamma}{2\pi}\sqrt{(H + H_K)(H + H_K + 4\pi M_s)} \qquad (1)$$

to extract the magnetic parameters of Py film. The magnetic parameters obtained from the fitting are, saturation magnetization ($M_s$) = 850 emu/cc, gyromagnetic ratio ($\gamma$) = 17.85 MHz/Oe and the anisotropy field ($H_K$) = 0. These will be further used for numerical micromagnetic simulations of the FMR spectra of the Py nano-cross structures.

The bias-field dependent FMR spectra (real part of $S_{11}$ parameter) for Py nano-cross arrays with 400 nm $\leq$ L $\leq$ 600 nm at different bias field orientations are shown in figures (2)-(4). They show rich SW properties, which vary non-monotonically with the bias field magnitude. These include the observation of a crossover between the two lowest frequency branches, followed by a sharp minima and maxima of the lowest frequency branch and merging of the two highest frequency branches followed by a Y-shaped mode splitting of the highest frequency branch with the decreasing bias field. These features are further modified by changing the nano-cross dimensions, i.e. the magnitude of the bias field at which these features appear, shift towards higher field values with the reduction of nano-cross arm length. Our purpose here is to study how these fascinating features are affected by the orientation of the bias magnetic field and whether new features can be generated.

The dip in the lowest frequency branch (Fig. 2(b)), which is a signature of mode softening appearing due to the variation in static magnetic configuration from an S-state to the onion-state, reduces significantly and shifts drastically to higher field value as the bias field angle $\phi$ is rotated by only 2° (Fig. 2(c)). This trend continues up to 5° (Fig. 2(d)), beyond which the dip disappears and instead a mode-crossover appears at around 300 Oe for $\phi$ = 10° for the nano-cross array with L = 600 nm. This continues up to 15°, beyond which the crossover also disappears and at $\phi$ = 45°, again a dip starts to appear. The bias field for crossover between the two lowest branches at around 700 Oe suddenly extends over a broad field range at $\phi$ = 10° with the appearance of a new SW branch and the crossover completely disappear at $\phi$ = 30°. The merging of the two highest frequency branches at around 500 Oe gets blurred at $\phi$ = 10° and it disappears at $\phi$ = 30°. The Y-shaped mode splitting of two highest frequency branches with positive and negative slopes appearing at around 150 Oe for $\phi$ = 0° shifts to higher bias field values with increasing values of $\phi$, which finally disappears at $\phi$ = 30°. With the variation of the nano-cross dimensions, some more changes are observed. For example, (a) the disappearance of the dip and consequent onset of cross-over for the sample with L = 500 nm still occurs at $\phi_a$ = 10° but the effect is rather feeble, while for L = 400 nm, the same occurs at $\phi_a$ = 5°. (b) The disappearance of Y-shaped mode splitting occurs at $\phi_b \approx$ 15° for L = 600 nm and it decreases gradually to $\phi_b \approx$ 10° for = 500 nm and $\phi_b \approx$ 5° for L = 400 nm. (c) The disappearance of mode crossover at higher field value, on the other hand, increases from $\phi_c \approx$ 10° at L = 600 nm to 15° and 30° for L = 500 nm and 400 nm, respectively. (d) On the contrary, the angle for appearance of mode splitting at higher field value for intermediate frequency branch remains unaltered with the arm length of the nano-cross. These observations are tabulated in Table (I). Here $\phi_a$, $\phi_b$, $\phi_c$ and $\phi_d$ correspond to the angle where features described as (a), (b), (c) and (d) above occur, respectively.

To interpret the experimental results, we performed micromagnetic simulations by using OOMMF [35] software. The arrays for performing simulation were mimicked from the SEM images and two-dimensional periodic boundary condition (2D-PBC) was applied for considering large areas of the arrays studied experimentally. The arrays were discretized into a number of rectangular prism-like cells of 4×4×20 nm$^3$ dimensions. The material parameters of the sample such as $\gamma$, $M_s$, and $H_K$ used in the simulations were extracted from the Kittel fit of the bias field dependent frequency of the Py thin film as discussed earlier, while the exchange stiffness constant ($A_{ex}$) = 1.3 × 10$^{-6}$ erg/cm is taken from literature [36]. The damping constant is used as $\alpha$ = 0.008 during dynamic simulations. The detailed methods of simulations are described elsewhere. Figures (2)-(4) show the simulated SW frequencies as function of magnetic field (unfilled symbols), which reproduced all the important features of the experimental results very well. We further simulated the power and phase profiles of the SW modes using a home built code [37]. Figure (5) shows simulated spatial distribution of phase profiles for various SW modes of the nano-cross array with L = 600 nm at three different bias fields (H) and seven different $\phi$ values. The power profiles of the same are shown in the supplementary figure (Fig. (1S)). SW modes with azimuthal character [32] do not show significant changes with $\phi$ due to azimuthal symmetry. The simulated spatial distribution of phase and power profiles corresponding to two highest frequency branches, which merge with the reduction of H, and intermediate frequency branch at H = 0.9 kOe for nano-cross with L = 600 nm are shown in Fig. (2S) and Fig. (3S), respectively (see supplementary material). The new mode, which appears from the splitting of highest frequency branch at higher H value, and intermediate frequency branch, which shows monotonic increase of resonance frequency with H, both show azimuthal character instead of DE character. We mainly focus here on radial SW modes. As $\phi$ changes, the phase profiles of the both SW modes of feature (a) remain same. These modes show a mixed backward volume (BV; n) and Damon Eshbach (DE; m)-like character with mode quantization numbers n = 11, m = 7 with opposite phases. Higher frequency branch of feature (a) disappears at $\phi \approx 15°$ for L = 600 nm. From phase profile the reason for disappearance of mode softening with the increasing of $\phi$ becomes clear. Mode softening is the result of the sudden switching of the magnetic configuration from S-state to onion state. Figure (6) shows a subtle increase in $\phi$ wipes away the onion state and instead S-state appears and stabilizes with further increase in $\phi$ as the bias field helps to align the spins accordingly. Consequently, as $\phi$ increases the height of minima due to mode softening decreases and the position of minima shifts to higher field value. For H $\approx$ 0.3 kOe at $\phi \geq 5°$, S-state of magnetic configuration starts to arise again. Beyond the critical angle ($\phi_a$), a mode crossover appears at the position of minima which is probably due to the reformation of the S-state. For feature (b) at $\phi = 0°$, the mode quantization numbers are n = 3, m = 1, which transforms to a new mode with n = 7, m = 3 at $\phi = 15°$ for H = 0.3 kOe. A new SW mode appears at the position of mode softening due to mode splitting at $\phi \approx 10°$ with mode quantization numbers n = 6, m = 4 for $\phi = 10°$, which change to n = 6, m = 6 for $\phi = 45°$. Higher frequency branch of feature (c) changes qualitatively with the variation of $\phi$. The mode quantization numbers of this mode get modified to n = 6, m = 5 at $\phi = 30°$ from n = 6, m = 1 at $\phi = 0°$. The mode quantization number of lower frequency branch of feature (b) are n = 5, m = 1 at $\phi = 0°$, which transforms to n = 5, m = 5 at $\phi = 45°$. The size of the nano-cross structure also has significant effect on the above-mentioned features (Table (I)). For feature (b) the $\phi_a$ value at which disappearance of mode softening and creation of new mode crossover occur, decreases with the reduction in L. This confirms that the bias field is more crucial for lower dimension of nano-cross in case of mode softening. The $\phi_a$ value related to disappearance of mode splitting at lower field decreases monotonically with the decrement in L. The competition between the spin configurations in the two orthogonal arms is the reason for appearance of this mode splitting.

As L decreases, the competition gets weaker as configurational anisotropy drops, which might be a possible reason for decreasing of $\phi_b$ with reduction in L. Interestingly, the $\phi_c$ values in feature (c) increases with the increment in L. The higher frequency branch between the two modes of this mode crossover is primarily showing BV-like mode character and it appears at H ≈ 0.7 kOe for L = 600 nm. This mode crossover shifts to higher H value as L decreases. This is responsible for the increment of $\phi_c$ with decreasing of L. At $\phi$ = 45°, the component of bias field in both the orthogonal arms become equal, which may be responsible for occurrence of minima in the lowest frequency mode at intermediate H value. A new mode splitting appears at higher H value for $\phi$ = 45°, which is independent of sizes of nano-cross in this size regime.

To understand the dynamics further, we have numerically calculated the magnetostatic field distributions in the nano-cross arrays and the corresponding contour plots are shown in Fig. 7(a) for four different orientations ($\phi$) of the bias field. The important observation here is that with the increase of $\phi$ value, the magnetic stray field line density inside the nano-cross structures increases which is probably due to increase of the uncompensated magnetic charges in the nano-cross structures in the direction of the bias field along both the arms. Line scans of the fields along the dashed lines are presented in Fig. 7(b), which reveals an important feature. With the increase of $\phi$, the internal fields decrease monotonically as plotted in Fig. 7(c). For $\phi$ = 0°, the maximum stray field value is $B_x$ ≈ 10.4 kOe which decreases to $B_x$ ≈ 7.6 kOe at $\phi$ = 45°. In particular, reduction of internal field is observed near the centre of the cross with increase in $\phi$. This feature is probably responsible for increase of $\phi_c$ with reduction in L. Figure 7(b) shows the width of the stray field distribution along x-axis at maximum $B_x$ value decreases with increase in $\phi$. The calculated magnetostatic field distributions indicate weak inter-cross stray magnetic fields. However, even such weak interaction field has some effect on the dynamics, particularly at lower bias field value. We performed FMR measurement for nano-cross arrays with arm-length (L) of 600 nm and with the edge-to-edge separation (S) 150 nm, 250 nm and 350 nm. Figure (4S) shows that the variation of S from 150 to 250 nm affects the minimum in the frequency for the lowest frequency branch due to mode softening, while no further changes occur as S is increased to 350 nm.

Figure (8) provides an exemplary demonstration of how the SW propagation direction can be manipulated using nano-cross array. To that end, using OOMMF software, we launched a time-varying field of "sinc" profile (frequency cutoff of 30 GHz) at the centre of the array over a small square region of 100 nm × 100 nm area. . We then simulated the propagation of the SW mode at f = 8.4 GHz for different orientation of the bias field. For $\phi = 0°$ the beam propagates along vertical direction, for $\phi = 90°$ the beam propagates along horizontal direction, for $\phi = 25°$ the beam propagates nearly isotropically in all directions while at $\phi = 45°$ the beam cease to propagate. This property may lead to possible application of the nano-cross array as a directional coupler or splitter. Figure (5S) in the supplementary material demonstartes a proposal of how densely packed nano-cross arrays may be used as frequency-dependent coupler. We performed further micromagnetic simulations to demonstrate possible application of nano-cross arrays as magnetic storage bits as shown in Fig. (6S) of the supplementary material. The above possible applications promote the ferromagnetic nano-cross arrays as building blocks of a variety of spintronic and magnonic devices.

## IV. CONCLUSIONS

In summary, we investigated the magnetization dynamics in $Ni_{80}Fe_{20}$ nano-cross arrays of varying sizes as a function of the orientation ($\phi$) of an external bias magnetic field using broadband ferromagnetic resonance technique. As bias field orientation ($\phi$) deviates from 0°, the chances of formation of onion state reduce. Consequently, the height of the dip in the lowest frequency branch reduces and dip position shifts to higher H value. Further increase of ϕ, causes a new mode crossover at lower H value in the place of the dip. Interestingly, the frequency difference between modes for mode crossover at higher H value increases with the increment of ϕ and further increase of ϕ leads towards disappearance of this high-field crossover resulting in two nearly parallel SW modes at ϕ = 30°, while at ϕ = 45° the higher frequency branch disappears. The intermediate frequency branch shows monotonic increase of frequency with bias field. The orientation of bias magnetic field (ϕ) strongly affects the Y-shaped mode splitting of highest frequency branch at lower H. The frequency gap between these two modes decreases with the increment of ϕ. Finally, at ϕ = $ϕ_b$ this mode splitting disappears. Higher field mode splitting shifts to lower H value with the enhancement of ϕ and finally disappears. The number of SW modes decreases at ϕ = 45° compared to ϕ = 30°. At ϕ = 45° the lowest frequency branch shows again a minimum at intermediate H. Simulated SW modes profile show that the modes showing strong dependence on bias field angle are of mixed BV-DE character. Interestingly, the DE mode quantization number m increases with ϕ for most SW modes. With the variation of the nano-cross dimension both quantitative and qualitative variations of the magnetization dynamics occur. Calculated magnetostatic field distributions reveal the origin of the variation in the SW mode frequencies and mode profiles. Such an ability to tune the spin configuration and magnetization dynamics in nano-cross structure by a subtle variation of external bias field angle is very important for the design of magnetic storage, memory, logic and magnonic devices as demonstrated by our further micromagnetic simulations.

## SUPPLEMENTARY MATERIAL

See supplementary material for (a) spatial power profiles of SW modes for L = 600 nm, (b) simulated spatial distribution of phase & power profiles corresponding to two highest frequency branches, which merge with the reduction of H and intermediate frequency branch at H = 0.9 kOe for nano-cross with L = 600 nm, (c) surface plots of bias field dependent SW mode frequencies for nano-cross arrays with arm-length (L) of 600 nm for varying edge-to-edge separation (S),(d) simulated amplitude profile of SW modes of different frequencies for densely packed nano-cross array when a time varying field of "sinc" profile excited locally at centre of the array, (e) simulated static magnetic configurations for nano-cross with L of 600 nm after applying different magnitudes of spin-polarized current.

## ACKNOWLEDGMENTS

The authors gratefully acknowledge the financial support from the Department of Science and Technology, Government of India under Grant No. SR/NM/NS-09/2011(G) and S. N. Bose National Centre for Basic Sciences, India (Grant No. SNB/AB/18-19/211).

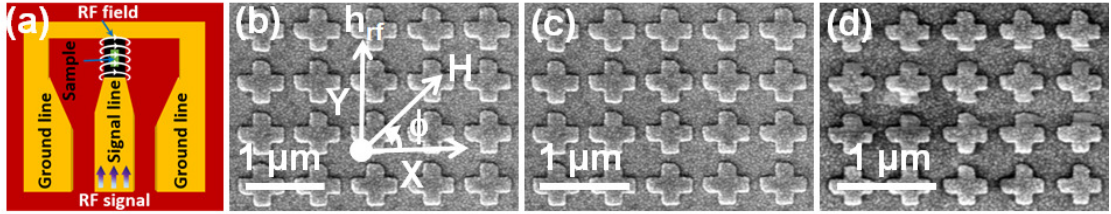

FIG. 1. (a) Schematic of our experimental geometry. (b)-(d) Scanning electron micrographs of Ni$_{80}$Fe$_{20}$ nano-cross arrays with varying arm lengths. The inset shows the orientation (φ) of external bias magnetic field.

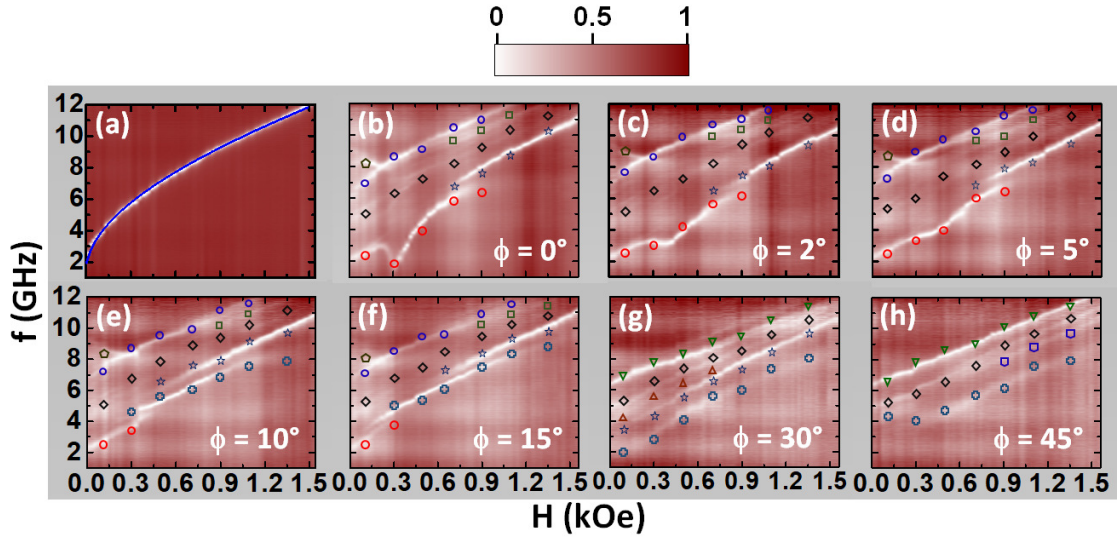

FIG. 2. Surface plots of bias field dependent SW mode frequencies for (a) thin film of 20 nm thickness and nano-cross with arm length (L) of 600 nm and for the bias field orientation (φ) of (b) 0°, (c) 2°, (d) 5°, (e) 10°, (f) 15°, (g) 30° and (f) 45°. (a) The Kittel fit is shown by the solid line. Simulated SW frequencies are shown by unfilled symbols. The color map is shown at the top of the figure.

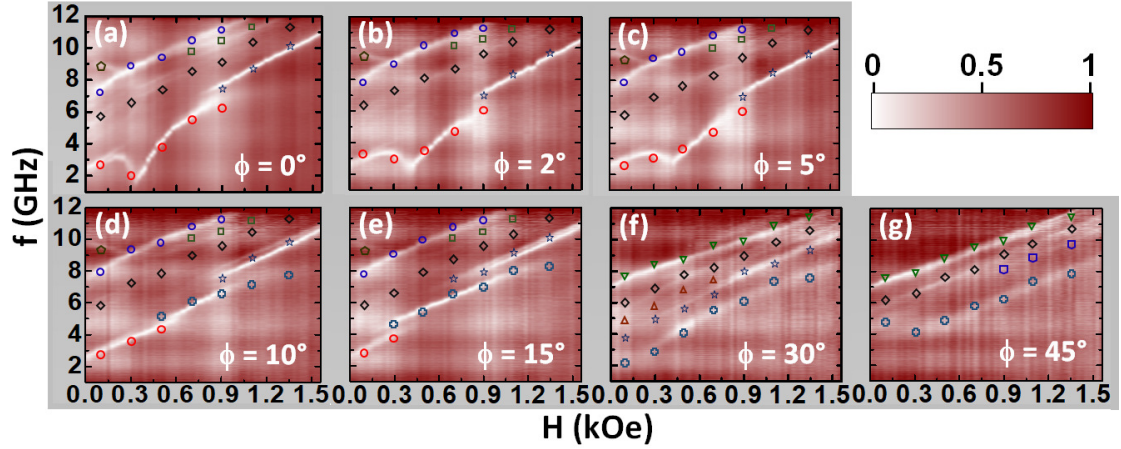

FIG. 3. Surface plots of bias field dependent SW mode frequencies for nano-cross arrays with arm length (L) of 500 nm and for the bias field orientation (ϕ) of (a) 0°, (b) 2°, (c) 5°, (d) 10°, (e) 15°, (f) 30° and (g) 45°. Simulated SW frequencies are shown by unfilled symbols. The color map is shown at the right side of the figure.

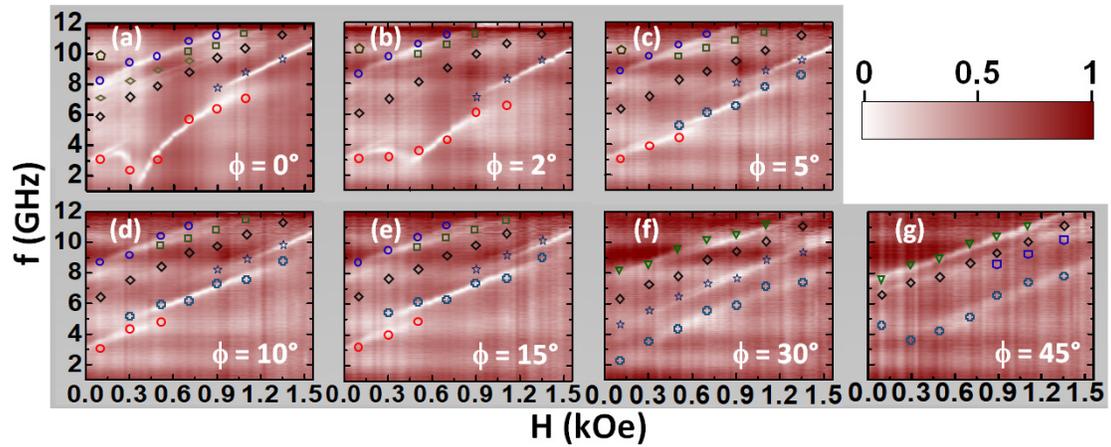

FIG. 4. Surface plots of bias field dependent SW mode frequencies for nano-cross arrays with arm length (L) of 400 nm and for the bias field orientation (ϕ) of (a) 0°, (b) 2°, (c) 5°, (d) 10°, (e) 15°, (f) 30° and (g) 45°. Simulated SW frequencies are shown by unfilled symbols. The color map is shown at the right side of the figure.

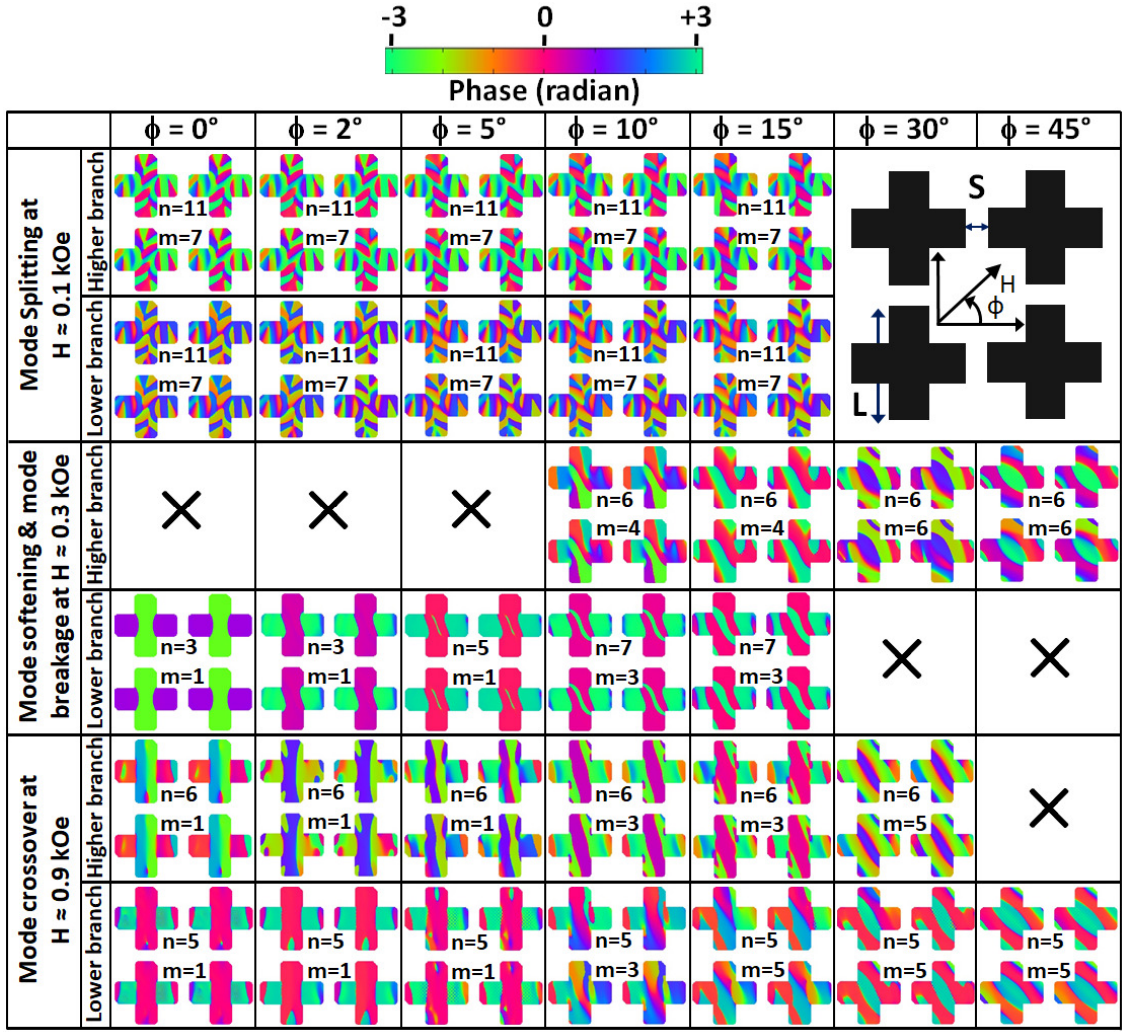

FIG. 5. Simulated spatial distribution of phase profiles corresponding to different salient SW modes at seven $\phi$ values and three different bias filed values for $Ni_{80}Fe_{20}$ nano-cross with L = 600 nm. The color map is shown at the top of the figure.

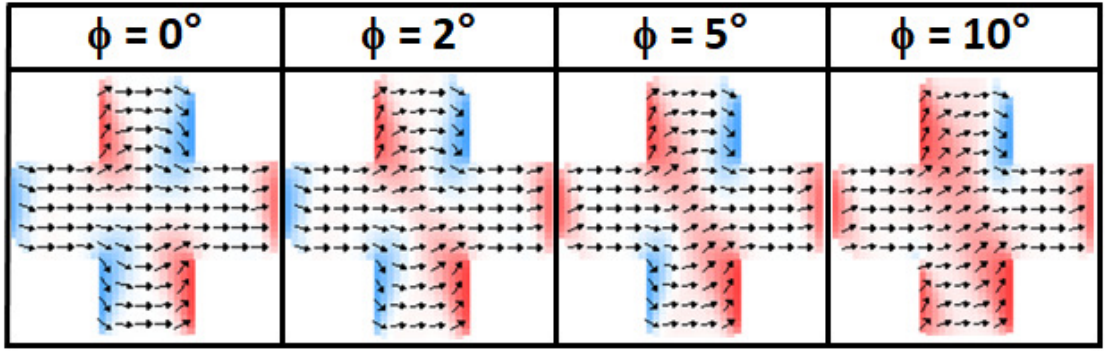

FIG. 6. Simulated static magnetic configurations for nano-cross with arm length (L) of 600 nm at bias field (H) = 0.5 kOe for four different ϕ values. We have shown a single cross from middle of the array to represent spin configurations prominently.

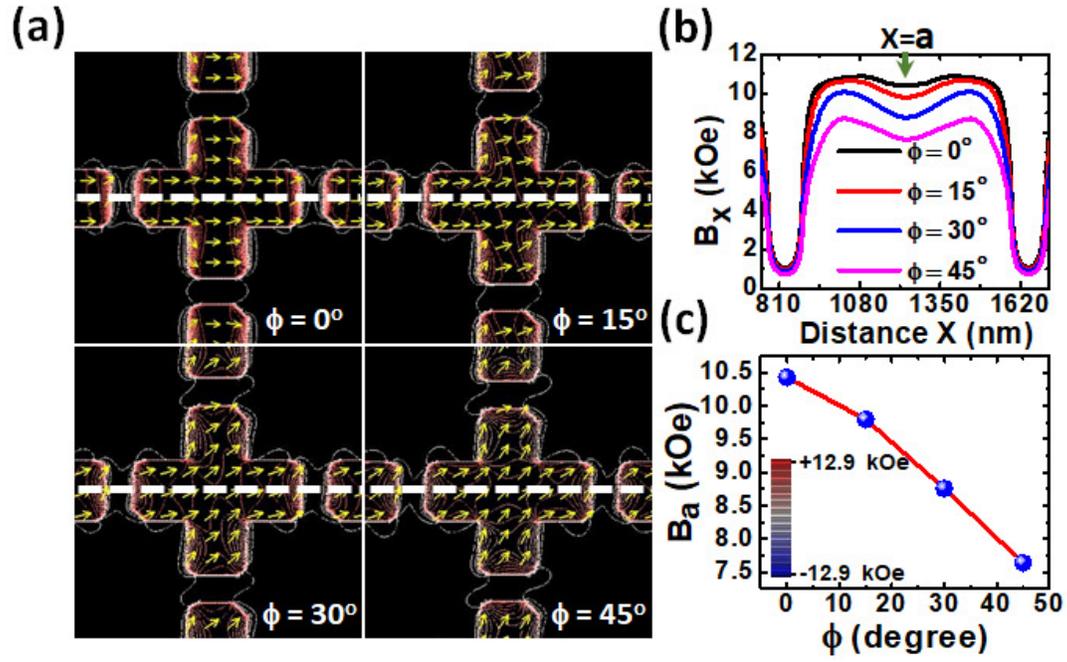

FIG. 7. (a) Contour plots of the simulated magnetostatic field distribution in $Ni_{80}Fe_{20}$ nano-cross arrays with arm length L = 600 nm in four different orientations (ϕ) of bias field for H = 0.6 kOe. Line scans are taken along the white dotted lines. (b) Line scans of the simulated magnetostatic fields. The color map is shown in the inset of the bottom right of the figure. (c) Effective magnetic field at the centre of the nano-cross for different ϕ values.

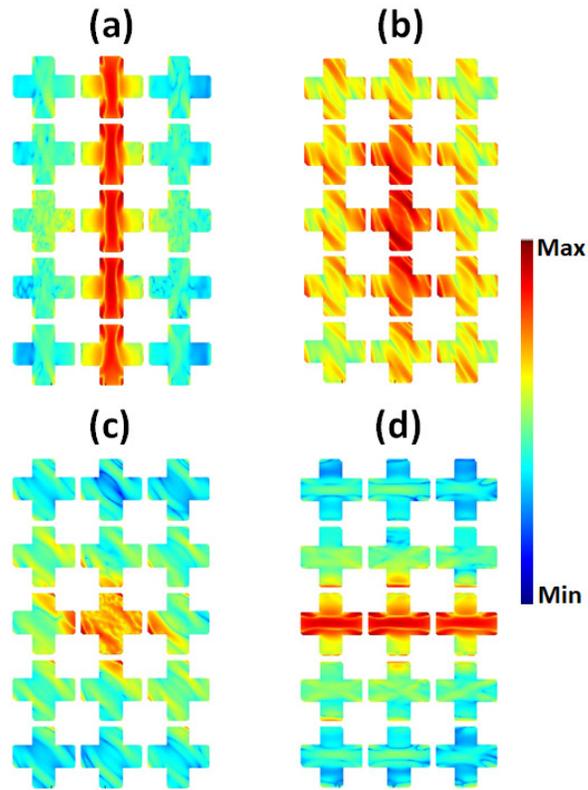

FIG. 8. Simulated power profiles of spin-wave modes of frequency (f) 8.4 GHz excited locally at centre of the array for (a) ϕ = 0°, (b) ϕ = 25°, (c) ϕ = 45° and (d) ϕ = 90° geometry.

Table I: Effect of varying the dimensions of nano-cross structure on critical angles of four prime features observed in the spin-wave dynamics.

| Arm length (L) | Disappearance of mode softening and creation of new mode crossover | Disappearance of Y-shaped mode splitting at lower H value | Disappearance of mode crossover at higher H value | Appearance of mode splitting at higher H value for intermediate frequency branch |
|---|---|---|---|---|
| 600 nm | $\phi_a \approx 10°$ | $\phi_b \approx 15°$ | $\phi_c \approx 10°$ | $\phi_d \approx 45°$ |
| 500 nm | $\phi_a \approx 10°$ | $\phi_b \approx 10°$ | $\phi_c \approx 15°$ | $\phi_d \approx 45°$ |
| 400 nm | $\phi_a \approx 5°$ | $\phi_b \approx 5°$ | $\phi_c \approx 30°$ | $\phi_d \approx 45°$ |